\documentclass[twocolumn]{revtex4}

\usepackage{amsmath}
\usepackage{graphicx}
\usepackage{color}

\newcommand{\D}{\mathrm{d}}

\newcommand{\e}{\mathrm{e}}

\newcommand{\eq}{Eq.}
\newcommand{\eqs}{Eqs.}
\newcommand{\fig}{Fig.}

\newcommand{\kbt}{k_{\mathrm{B}}T}

\newcommand{\lb}{l_\mathrm{B}}
\newcommand{\ld}{\lambda_{_{\mathrm{D}}}}
\newcommand{\kd}{\kappa_{_\mathrm{D}}}

\begin{document}

\title{Interaction between Heterogeneously Charged Surfaces: Surface Patches and Charge Modulation}

\author{Dan Ben-Yaakov, David Andelman}
\email{andelman@post.tau.ac.il}
\affiliation{Raymond \& Beverly Sackler School of Physics and Astronomy\\ Tel Aviv
University, Ramat Aviv, Tel Aviv 69978, Israel}

\author{Haim Diamant}
\affiliation{Raymond \& Beverly Sackler School of Chemistry\\ Tel Aviv University, Ramat Aviv, Tel Aviv 69978, Israel}

\date{resubmitted to PRE -- 02/09/2012}

\begin{abstract}
When solid surfaces are immersed in aqueous solutions, some of their charges can dissociate and leave behind
charge patches on the surface. Although the charges are distributed heterogeneously
on the surface, most of the theoretical models treat them as homogeneous.
For overall non-neutral surfaces, the assumption of surface charge homogeneity is rather reasonable, since the leading terms
of two such interacting surfaces depend on the non-zero average charge. However,
for overall neutral surfaces, the nature of the surface charge distribution
is crucial in determining the inter-surface interaction. In the present work
we study the interaction between two charged surfaces across an aqueous solution for several charge distributions.
The analysis is preformed within the framework of the linearized Poisson-Boltzmann theory. For periodic charge distributions the interaction is found to be repulsive at small separations, unless the two surface distributions
are completely out-of-phase with respect to each other.  For quenched random charge distributions we find
that due to the presence of the ionic solution in between the surfaces, the
inter-surface repulsion dominates over the attraction in the linear regime of the Poisson-Boltzmann theory.
The effect of
quenched charge heterogeneity is found to be particularly substantial in the case of large charge domains.
\end{abstract}

\maketitle

\section{Introduction\label{sec:Introduction}}

\begin{figure}
\begin{centering}
\includegraphics[scale=0.3]{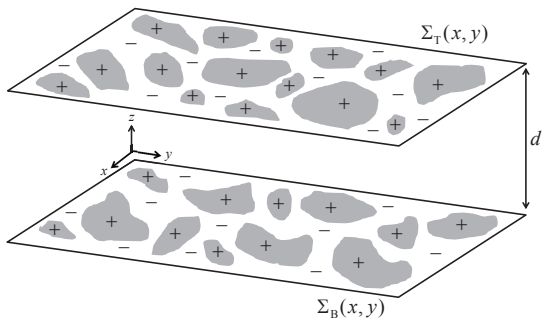}
\par\end{centering}
\caption{Schematic drawing of two planar heterogeneously charged surfaces immersed in ionic solution. The gray
regions are positively charged and the white ones are negatively
charged. The charge distributions of the bottom and top surfaces are given by $\Sigma_\mathrm{B}\left(x,y\right)$ and $\Sigma_\mathrm{T}\left(x,y\right)$, respectively. The inter-surface separation is $d$.\label{fig:Schematic}}
\end{figure}

Long-range interactions between charged surfaces substantially influence the structural
properties of soft materials, such as lipid membranes and colloidal suspensions
\cite{israel_book,safran_book,evans_book}. The nature of the long-range interaction between two charged surfaces immersed in an ionic solution is mainly determined by electrostatic interactions,
mediated by the ionic solutes and polar solvent. When modeling the electrostatic
interaction, surfaces are usually assumed to be homogeneously
charged~\cite{verwey,israel_book,andelman,benyaakov2007}. However, most charged surfaces
in soft matter are heterogeneous over a certain length scale. A schematic drawing of two such heterogeneously charged surfaces is presented in \fig~\ref{fig:Schematic}. While the interaction between two homogeneous surfaces depends on the average
surface charge and the inter-surface separation, the interaction between two
heterogeneous surfaces depends on the intra- and inter-surface charge correlations as well.

Several experimental studies \cite{exp2001,perkin2005,meyer2005,zhang2005,perkin2006,meyer2006,ball2008,hammer2010,silbert2011,drelich2011,popa2010} measured inter-surface forces for different configurations
of surface charge heterogeneities. For example, it was demonstrated \cite{exp2001,perkin2005,meyer2005,zhang2005,perkin2006,meyer2006,ball2008,hammer2010,silbert2011} that coating a negatively charged mica surface
with a cationic surfactant monolayer may lead to the formation of positively charged
bilayer patches, which neutralize the negative patches of the bare mica. A
long-range attractive force was measured \cite{exp2001,perkin2005,meyer2005,zhang2005,perkin2006,meyer2006,ball2008,hammer2010,silbert2011} between two such coated mica surfaces. In another setup, it was found that adsorption of positive
polyelectrolytes on negatively-charged spherical beads can lead to
charge inversion of the beads \cite{drelich2011,popa2010}. The adsorbed
polyelectrolyte forms positively charge domains, while patches of bare
regions on the beads remain negatively charged. It has been shown
\cite{popa2010} that by controlling the amount of adsorbed polyelectrolyte, overall neutral particles
with a patchy heterogeneous charge distribution can be prepared. In this
special case of neutral beads, the inter-particle interaction is attractive,
while for smaller or larger amount of adsorbed polyelectrolyte (where
the average bead charge is, respectively, negative or positive) the interaction is
repulsive.

The interaction between two {\it periodically ordered} charged surfaces has been studied theoretically in several works \cite{richmond1974,richmond1975,muller1983,discrete1992,miklavic1994,holt1997,khachatourian1998,stankovitch1999,velegol2001} within the framework of the linearized Poisson-Boltzmann (PB) equation. For non-neutral surfaces, the leading interaction term depends on the average charge. For neutral surfaces, where the average charge is zero, the interaction depends strongly on the relative phase between the two charge distributions, and can vary from being repulsive to attractive. For two surfaces with identical average charge and small amplitude
charge modulation it was shown, in the non-linear regime, that the repulsive interaction is weaker than the repulsion between two uniformly charged surfaces with the same average charge~\cite{miklavciv1995,white2002,lukatsky2002a,lukatsky2002b,fleck2005,landy2010}. However, for  non-zero average charge, the charge modulation leads only to a small correction, as compared to the leading interaction term determined by the average charge.
%

In a  more general case of experimental relevance the surface charges are distributed {\it randomly} (and not periodically).
We further make the important distinction between quenched and annealed cases of surface charge disorder.
In the \emph{annealed} case, the surface charges are in thermodynamic equilibrium with
other system variables, such as inter-surface separation, temperature and concentration of other species.
The attraction can then be caused by self-adjusting of surface charge domains, where positively
charge patches on one surface position themselves against negatively charge patches on the second surface and vice versa \cite{naydenov2007,brewster2008,jho2011,deLaCruz2006}. For this case to hold, the typical time-scale of patch rearrangement must be substantially shorter than the measurement time-scale, such that the self-adjustment (annealing) of charged
domains on the surfaces has enough time to be completed.

In the \emph{quenched} case, the surface charges are frozen and independent of other system variables. For a quenched distribution of surface charges with no inter- and intra-surface correlations, it was reported that such  heterogeneities have no effect on the two-surface interaction at the mean-field level \cite{naji2005,naji2008}, and the interaction depends solely on the average surface charge.

Another interesting case of quenched disorder, to be considered in detail in the present work, is the case where finite-size charge domains (patches) are randomly distributed on each surface.
In some of the experiments such random charge domains stem from the specific surface preparation
\cite{exp2001,perkin2005,meyer2005,zhang2005,perkin2006,meyer2006,
ball2008,hammer2010,silbert2011}.

The outline of our paper is as follows. The model is formulated
in Sec.~\ref{sec:The-Model}. In Sec.~\ref{sub:Quenched-disorder}
the interaction between surfaces with periodic charge modulations is revisited
and a crossover from attraction to repulsion is discussed.
Quenched random charge distributions are treated in Sec.~\ref{sub:Random-quenched-disorder}
for several cases within the linear regime of the PB equation. In particular, the limit of small domains having a size in the molecular range is compared with the limit of large domains.
In Sec.~\ref{sec:discussion} we discuss the implications of our results, and concluding remarks are
presented in Sec.~VI.
Finally, in the limit of infinitely large charge domains,
we compare in the Appendix the predictions of the linear regime to those of the non-linear one considered in
Ref.~\cite{silbert2011}, where an attractive inter-surface interaction was predicted
in the limit of infinitely-large charge domains. An interesting crossover is found
from inter-surface attraction of strongly charged surfaces to repulsion for weakly charged surfaces, emphasizing
even further the role of the domain lateral size.

\section{The Linear Poisson-Boltzmann Model\label{sec:The-Model}}

We consider an aqueous solution as depicted in Fig.~\ref{fig:Schematic}, which is
confined between two semi-infinite
and planar charged surfaces, located at $z=\pm d/2$, where the $\hat{z}$-axis
is perpendicular to the surfaces. The surface-charge density distribution
(charge per unit area) is given
by $\Sigma_\mathrm{B}\left(x,y\right)$ for the bottom surface at $z=-{d}/{2}$,
and $\Sigma_\mathrm{T}\left(x,y\right)$
for the top one at $z={d}/{2}$.
As we are interested in the effect of \emph{quenched} charge heterogeneity,
we model the lateral surface charge distributions $\Sigma_\mathrm{B}$
and $\Sigma_\mathrm{T}$ as fixed charge boundary conditions. Namely, they are frozen
and independent of other system variables.

The aqueous solution contains a  1:1 monovalent salt ions.
The solvent (water) is modeled as a homogeneous dielectric background
with dielectric constant $\varepsilon=80$. The solution is coupled
to a reservoir of ionic density $n_{b}$. The electrostatic potential,
$\psi$, and the ionic densities, $n_{\pm}=n_b \e^{\mp e \psi/\kbt}$, are calculated via the
Poisson-Boltzmann (PB) equation:
\begin{equation}
\nabla^{2}\psi=\frac{8\pi en_{b}}{\varepsilon}\sinh\left(e\psi/\kbt\right)\,,\label{eq:nl_PB}
\end{equation}
where $e$ is the electron charge, $k_{\mathrm{B}}T$
is the thermal energy, and $k_\mathrm{B}$ is the Boltzmann constant.
By rescaling the potential $\psi\rightarrow\phi\equiv e\psi/\kbt$ the PB equation is rewritten for $\phi$,
the dimensionless electrostatic potential as:
\begin{equation}
\nabla^{2}\phi=\kd^2\sinh\phi\,,\label{eq:d_nl_PB}
\end{equation}
where $\kd=\ld^{-1}=\sqrt{8\pi l_{\mathrm{B}}n_{b}}$ is
the inverse Debye screening length and $l_\mathrm{B}=e^2/\left(\varepsilon\kbt\right)$
is the Bjerrum length. The boundary conditions at $z=\pm d/2$ are given
by:
\begin{equation}\nonumber
\left.\frac{\partial\phi}{\partial z}\right|_{-d/2}=-\sigma\left(x,y\right)\,,\label{eq:bc_1}
\end{equation}
\begin{equation}
\left.\frac{\partial\phi}{\partial z}\right|_{d/2}=\eta\left(x,y\right),\label{eq:bc_2}
\end{equation}
where $\sigma\equiv4\pi l_{\mathrm{B}}\left({\Sigma_\mathrm{B}}/{e}\right)$
and $\eta\equiv4\pi l_{\mathrm{B}}\left({\Sigma_\mathrm{T}}/{e}\right)$
are the two rescaled surface charge densities, having dimensions of inverse length. {For homogeneous surface charge densities the rescaled densities $\sigma$ and $\eta$ are the inverse of the Gouy-Chapman length \cite{safran_book} (up to a factor of two)
that characterizes the thickness of the condensed ionic layer near a homogeneously charged surface.}

The interaction energy between the two surfaces as a function of the
separation $d$ is given by the free energy, $F\left(d\right)$,
subject to the boundary conditions, \eq~(\ref{eq:bc_2}).
We obtain $F\left(d\right)$ by using the charging method
\cite{verwey}, and calculate the work needed
to increase the surface charge incrementally, at each point on the surface, from zero to the desired
final value. The charging free-energy of two surfaces coupled with an ionic reservoir
of density $n_b$ is given by:
\begin{align}
\left(\frac{4\pi\kd l_{\mathrm{B}}}{k_{\mathrm{B}}T}\right)F\left(d\right)=
\quad\quad\quad\quad\quad\quad\quad\quad\quad\quad\quad\quad\quad\nonumber\\ \kd\iint\mathrm{d}x\mathrm{d}y\,\int_{\left(0,0\right)}^{\left(\sigma,0\right)}
\mathrm{d}\sigma'\,\phi\left[\sigma',\eta'=0\right]_{z=-d/2}\nonumber \\
 + ~\kd\iint\mathrm{d}x\mathrm{d}y\,\int_{\left(\sigma,0\right)}^{\left(\sigma,\eta\right)}
\mathrm{d}\eta'\,\phi\left[\sigma,\eta'\right]_{z=d/2}\,,\label{eq:free_energy}
\end{align}
where throughout the paper $F\left(d\right)$ is made dimensionless by rescaling it,
$F\to ({4\pi\kd l_{\mathrm{B}}}/k_{\mathrm{B}}T)F$. The electrostatic potential on the surfaces, $\phi$,
is written in the above equation as a function of the distributions $\sigma$ and $\eta$ to stress that the integration is performed with respect to these variables.

Equations (1)-(4) are valid for the general PB theory, while hereafter we concentrate on
the linear PB equation, being valid
as long as $\phi=e\psi/\kbt\ll1$ ($\psi\ll25$mV):
\begin{equation}
\nabla^{2}\phi=\kd^2\phi\,.\label{eq:dim_l_PB}
\end{equation}
The in-plane Fourier transform of $\phi\left(\boldsymbol{\rho},z\right)$
in the $\boldsymbol{\rho}=\left(x,y\right)$ plane is denoted by ${\phi}_{\,\mathbf{k}}\left(z\right)$
and is given by:
\begin{equation}
{\phi}_{\,\mathbf{k}}\left(z\right)=\int\mathrm{d}^{2}\rho\,\mathrm{e}^{i\mathbf{k}\cdot\boldsymbol{\rho}}
\phi\left(\boldsymbol{\rho},z\right)\,,\label{eq:FT_phi}
\end{equation}
where $\mathbf{k}$ is the inplane wavevector. Using ${\phi}_{\,\mathbf{k}}$ we get from \eq~(\ref{eq:dim_l_PB}):
\begin{eqnarray}
\frac{\partial^{2}{\phi}_{\,\mathbf{k}}}{\partial z^{2}}&=&q^{2}{\phi}_{\,\mathbf{k}}\nonumber\,,\\
\label{eq:FT_l_PB}
q&=&\sqrt{\kd^2+k^{2}}\,.
\end{eqnarray}
The boundary conditions for ${\phi}_{\,\mathbf{k}}$
are similarly obtained by Fourier transforming \eq~(\ref{eq:bc_2}):
\begin{equation}\nonumber
\left.\frac{\partial{\phi}_{\,\mathbf{k}}}{\partial z}\right|_{-d/2}=-{\sigma}_{\mathbf{k}}\,,\label{eq:FT_bc_1}
\end{equation}
\begin{equation}
\left.\frac{\partial{\phi}_{\,\mathbf{k}}}{\partial z}\right|_{d/2}={\eta}_{\mathbf{k}}\,,\label{eq:FT_bc_2}
\end{equation}
where ${\sigma}_{\mathbf{k}}$, ${\eta}_{\mathbf{k}}$ are, respectively, the Fourier
transforms of $\sigma\left(\boldsymbol{\rho}\right)$ and $\eta\left(\boldsymbol{\rho}\right)$:
\begin{equation}\nonumber
{\sigma}_{\mathbf{k}}=\int\mathrm{d^{2}}{\rho}\,
\mathrm{e}^{i\mathbf{k}\cdot\boldsymbol{\rho}}\sigma\left(\boldsymbol{\rho}\right)\,,
\end{equation}
\begin{equation}
{\eta}_{\mathbf{k}}=\int\mathrm{d}^{2}{\rho}\,
\mathrm{e}^{i\mathbf{k}\cdot\boldsymbol{\rho}}\eta\left(\boldsymbol{\rho}\right)\,.
\end{equation}
Solving for the $z$-dependence of ${\phi}_{\,\mathbf{k}}$, Eqs.~(\ref{eq:FT_l_PB})-(\ref{eq:FT_bc_1}) yields:
\begin{equation}
{\phi}_{\,\mathbf{k}}\left(z\right)=\frac{\left({\sigma}_{\mathbf{k}}+{\eta}_{\mathbf{k}}\right)
\cosh\left(qz\right)}{2q\sinh\left(qd/2\right)}+\frac{\left({\eta}_{\mathbf{k}}
-{\sigma}_{\mathbf{k}}\right)\sinh\left(qz\right)}{2q\cosh\left(qd/2\right)}\,.\label{eq:sol_FT_phi}
\end{equation}
The free energy can be re-expressed in terms of the Fourier transforms ${\sigma}_{\mathbf{k}}$,
${\eta}_{\mathbf{k}}$ and ${\phi}_{\,\mathbf{k}}$. By substituting
\eq~(\ref{eq:sol_FT_phi}) into the free energy, \eq~(\ref{eq:free_energy}), we obtain:
\begin{align}
 F\left(d\right)=
\quad\quad\quad\quad\quad\quad\quad\quad\quad\quad\quad\quad\quad\quad\quad\quad\quad\quad\quad\quad
\nonumber
\\
\kd\int\frac{\mathrm{d}^{2}k}{\left(2\pi\right)^{2}}\,
\frac{{\sigma}_{\mathbf{k}}{\eta}_{-\mathbf{k}}+{\sigma}_{-\mathbf{k}}{\eta}_{\mathbf{k}}
+\left({\sigma}_{\mathbf{k}}{\sigma}_{-\mathbf{k}}+{\eta}_{\mathbf{k}}{\eta}_{-\mathbf{k}}\right)\mathrm{e}^{-qd}}{2q\sinh\left(qd\right)}
\,,
\label{eq:dim_free_energy_2}
\end{align}
where the reference contribution at $d\rightarrow\infty$ was subtracted
in order to obtain zero interaction between surfaces at infinite separation,
$F(d\rightarrow \infty)=0$. Because of the linearity, the PB free energy can also be written as a sum of two decoupled variables ${\sigma}_{\mathbf{k}}^{\pm}\equiv \left({\sigma}_{\mathbf{k}}\pm{\eta}_{\mathbf{k}}\right)/2$:
\begin{align}\nonumber
 F\left(d\right)
=\kd\int\frac{\mathrm{d}^{2}k}{\left(2\pi\right)^{2}}
\,\left[\frac{\mathrm{e}^{-qd/2}}{q\sinh\left(qd/2\right)}{\sigma}_{\mathbf{k}}^{+}{\sigma}_{-\mathbf{k}}^{+}
\right.\quad \quad \\  \left.-\frac{\mathrm{e}^{-qd/2}}{q\cosh\left(qd/2\right)}{\sigma}_{\mathbf{k}}^{-}{\sigma}_{-\mathbf{k}}^{-}\right]
\,.\label{eq:decoupled_free_energy}
\end{align}
The first term is positive definite and decreases monotonically as a function
of the separation $d$, contributing to the repulsive part of the inter-surface interaction, while
the second term, being negative definite and a monotonic increasing
function of $d$, gives the attractive part of the interaction.

Quite generally, it can be stated that the
magnitude of the repulsive term is larger than the attractive one
for all values of $q$ and $d$, because of the inequality
$\sinh\left(qd/2\right)<\cosh\left(qd/2\right)$.
Furthermore, for $d\rightarrow0$ the magnitude of the repulsive
term diverges, while the attractive term approaches a finite limiting
value. Consequently, only when ${\sigma}_{\mathbf{k}}^{+}{\sigma}_{-\mathbf{k}}^{+}=0$
for any wavenumber $\mathbf{k}$, does the interaction become attractive for {\it any} $d$.
This occurs only when the distributions of the two surfaces
are completely out-of-phase (anti-symmetric) with each other, $\sigma\left(\boldsymbol{\rho}\right)=-\eta\left(\boldsymbol{\rho}\right)$. Furthermore,
by properly choosing $\sigma_\mathbf{k}^\pm$, $F(d)$ can be made attractive for large enough $d$.

\section{Periodic Surface Charge Modulation\label{sub:Quenched-disorder}}

In the previous section the distributions $\sigma\left(\boldsymbol{\rho}\right)$ and
$\eta\left(\boldsymbol{\rho}\right)$ have been taken to be arbitrary. Hence, the inter-surface interaction of \eqs~(\ref{eq:dim_free_energy_2})-(\ref{eq:decoupled_free_energy})  applies to any form of surface charge distributions.
We treat next two interacting charged surfaces each characterized by a periodic charge modulation.
As periodic charge modulations have been investigated
previously~\cite{richmond1974,richmond1975,muller1983,discrete1992,miklavic1994,holt1997,khachatourian1998,stankovitch1999,velegol2001},
we revisit the periodic case in order to discuss the interesting
crossover from attraction to repulsion and the limiting behavior for small and large periodic domains. It is also instructive to compare the periodic case with the quenched disorder that will be derived later in Sec.~IV.

\subsection{Single Mode\label{sub:Single-mode-modulation}}

\begin{figure*}
\begin{centering}
\includegraphics[width=0.38\textwidth]{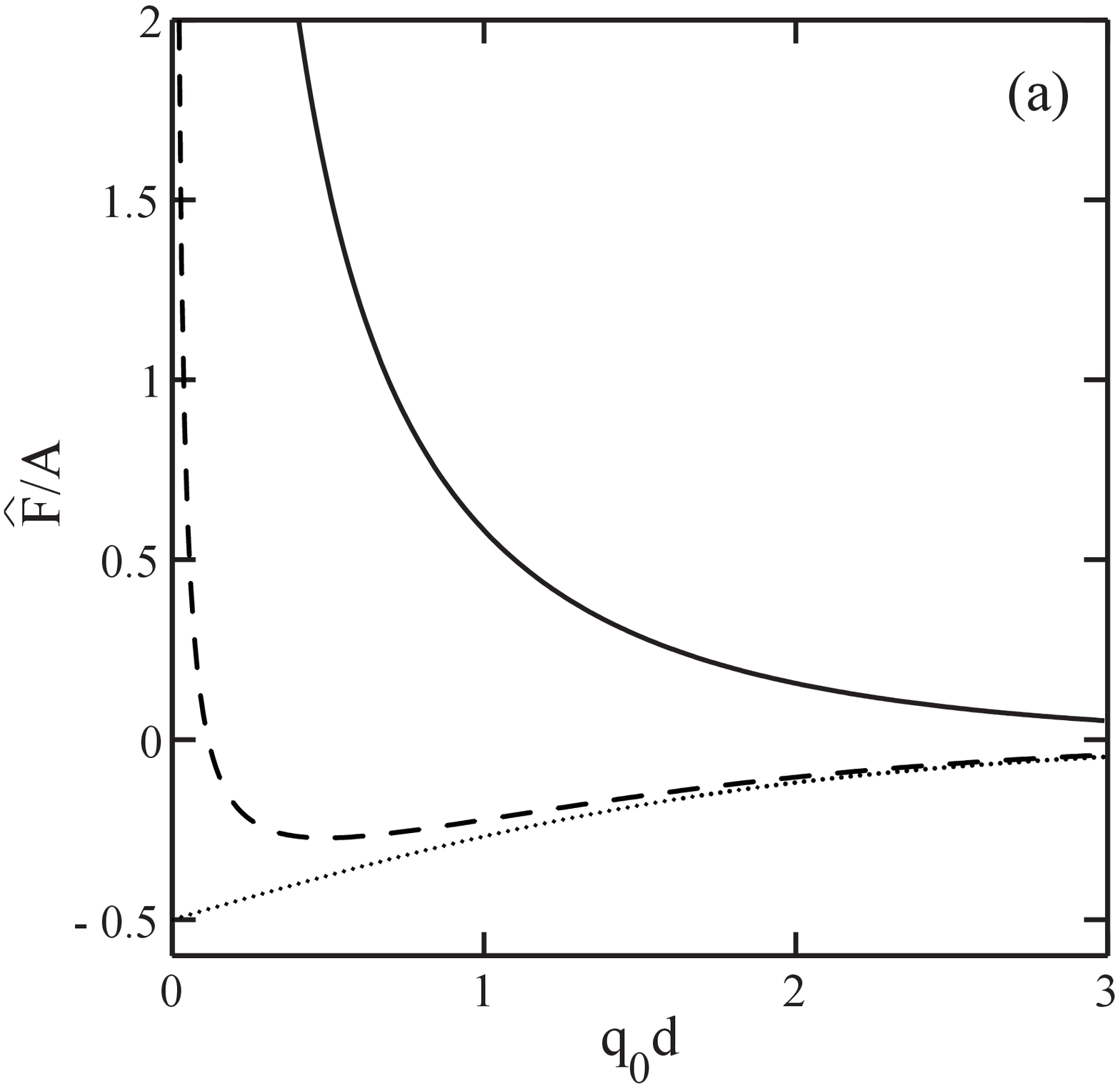}~~~~~~~\includegraphics[width=0.38\textwidth]{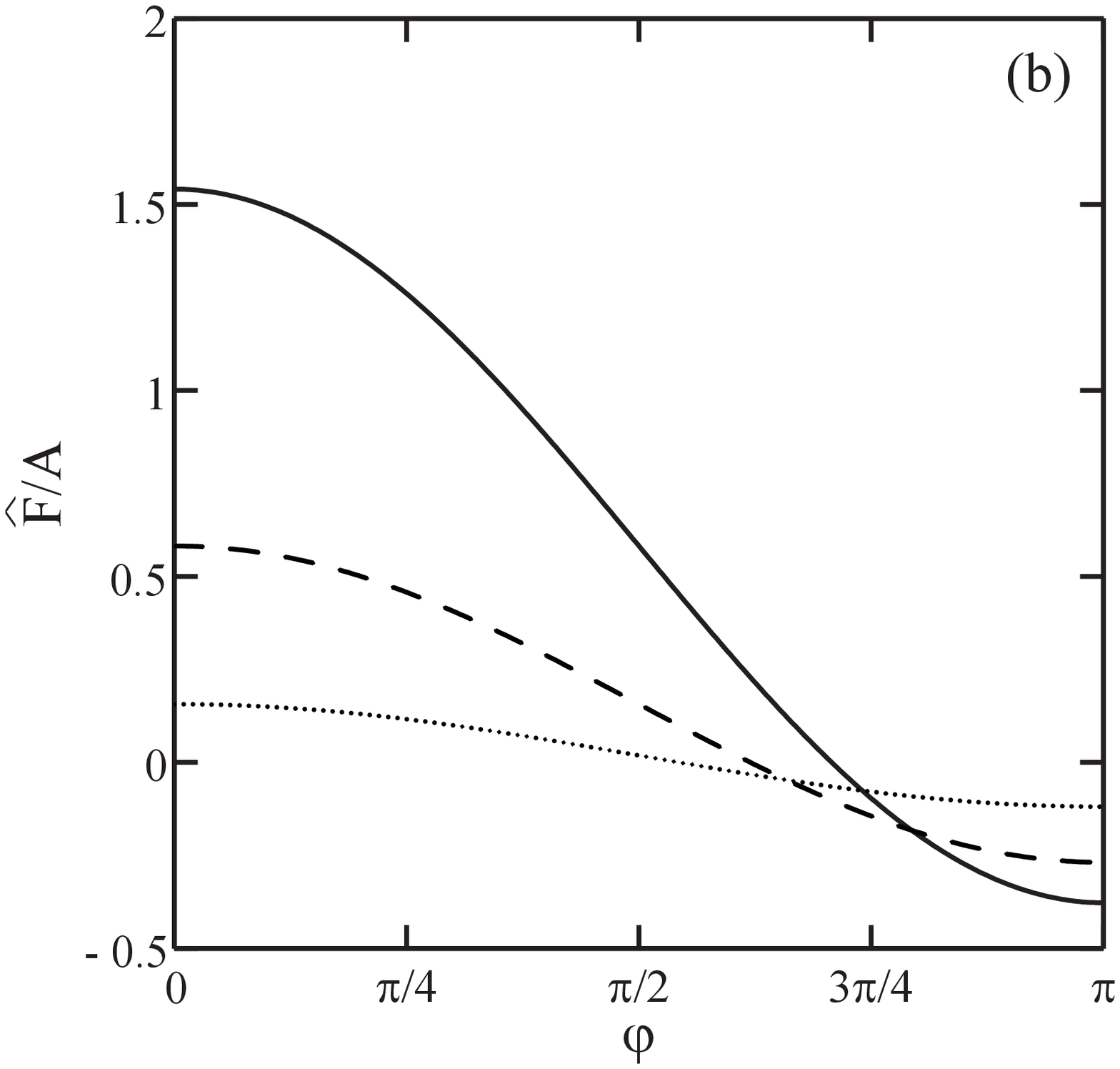}
\par\end{centering}
\caption{\label{fig:The-interaction-energy.}The inter-surface interaction (per unit area) $\widehat{F}/{A}$
for a single $q_0$-mode, rescaled by the prefactor of \eq~(\ref{eq:single_mode_energy_2}),
$\widehat{F}=\left( q_0/\kd C_+^2\right)F$, and plotted
for the equal amplitude case: $C_\sigma=C_\eta=C_{+}$ ($C_{-}=0$).
(a) The inter-surface interaction as a function of the inter-surface
separation $d$ in units of $q_0^{-1}$. The values of the relative angle $\varphi$ are 0, $0.85\pi$, $\pi$ for
the solid, dashed and dotted lines, respectively. (b) The inter-surface interaction as a function of the relative phase $\varphi$. The values
of $q_{0}d$ are 0.5, 1 and 2 for the solid, dashed and dotted lines,
respectively.}
\end{figure*}

We first consider two surfaces characterized
by the same single-$\mathbf{k}_{0}$ mode modulation
\begin{equation}\nonumber
\sigma\left(\boldsymbol{\rho}\right)=C_{\sigma}\cos\left(\mathbf{k}_{0}\cdot\boldsymbol{\rho}\right)
\,,\label{eq:single_mode_sigma_1}
\end{equation}
\begin{equation}
\eta\left(\boldsymbol{\rho}\right)
=C_{\eta}\cos\left(\mathbf{k}_{0}\cdot\boldsymbol{\rho}+\varphi\right)
\,,\label{eq:single_mode_sigma_2}
\end{equation}
where $C_{\sigma}$ and $C_{\eta}$ (having units of inverse length) are the modulation amplitudes of the two surfaces, taken to be positive without loss of generality. The angle $-\pi\leq\varphi\leq\pi$ is the relative
phase between the two charge modulations. It reflects an in-phase arrangement
of like charges at $\varphi=0$, and an anti-phase one at
$\varphi=\pm\pi$.

The inter-surface interaction per unit area, $ F\left(d\right)/A$, is given by:
\begin{equation}
\frac{ F\left(d\right)}{A}=
\frac{2\kd C_{\sigma}C_{\eta}\cos\varphi+\kd\mathrm{e}^{-q_{0}d}
\left(C_{\sigma}^{2}+C_{\eta}^{2}\right)}{4 q_{0}\sinh(q_{0}d)}
\,,\label{eq:single_mode_energy}
\end{equation}
where $A$ is the lateral surface area and $q_{0}=\sqrt{\kd^2+k_{0}^{2}}$.
For small separations, $q_{0}d\ll1$, the energy scales as
$ F/A\sim\kd\left(q_{0}^{2}d\right)^{-1}$,
while for large separations, $q_{0}d\gg1$, the energy decays exponentially,
$ F/A\sim \kd q_{0}^{-1}\mathrm{e}^{-q_{0}d}$.

Rewriting $ F/A$ in terms of $C_{\pm}\equiv \left(C_{\sigma}\pm C_{\eta}\right)/2$
as in Eq.~(\ref{eq:decoupled_free_energy}) leads to:
\begin{align}
\frac{ F\left(d\right)}{A}=
\frac{\kd}{2 q_{0}}\frac{\mathrm{e}^{-q_{0}d}+\cos\varphi}{\sinh\left( q_{0}d\right)}\,C_{+}^{2}
\quad
\nonumber
\\
+\frac{\kd}{2 q_{0}}\frac{\mathrm{e}^{-q_{0}d}-\cos\varphi}{\sinh\left( q_{0}d\right)}\,C_{-}^{2}
\,.\label{eq:single_mode_energy_2}
\end{align}
Here both the $C_{+}^2$ and $C_{-}^2$ terms can be either repulsive or attractive, depending
on the values of $d$ and $\varphi$.

In \fig~\ref{fig:The-interaction-energy.} the inter-surface interaction per unit area
${ F\left(d\right)/A}$ is presented for charge modulations
with two equal amplitudes, $C_{\sigma}=C_{\eta}=C_{+}$ .
The second term of \eq~(\ref{eq:single_mode_energy_2})
then vanishes  as $C_{-}=0$, and the sign of the interaction energy is determined by
the relative phase $\varphi$ and $d$. The repulsion is maximal with respect
to $\varphi$ for the in-phase state $\varphi=0$, while for the anti-phase
state, $\varphi=\pm\pi$, the attraction is maximal:
\begin{align}
\frac{ F\left(d\right)}{A}=
\begin{cases}
{\frac{\kd}{2q_{0}}\frac{\mathrm{e}^{-q_{0}d/2}}{\sinh(q_{0}d/2)}}C_{+}^{2} ~~~~~~ \mathrm{for\,\,\,\,}\varphi=0\\
~~&~~\\
-\frac{\kd}{2q_{0}}\frac{\mathrm{e}^{-q_{0}d/2}}{\cosh(q_{0}d/2)}C_{+}^{2} ~~~ \mathrm{for\,\,\,\,}\varphi=\pm\pi
\end{cases}
\end{align}
Furthermore, for modulated surfaces with equal amplitudes, $C_{\sigma}=C_{\eta}$, the interaction in Eq.~(\ref{eq:single_mode_energy_2})
is purely repulsive in the range $\left|\varphi\right|<\pi/2$, where the cosine is positive. When $\cos\varphi$ becomes negative,
in the range $\pi/2<\left|\varphi\right|<\pi$, the inter-surface interaction
$ F\left(d\right)$ has a minimum at $d_{c}=q_{0}^{-1}\cosh^{-1}\left(1/\left|\cos\varphi\right|\right)$,
leading to a crossover from attraction for $d>d_{c}$ to repulsion
for $d<d_{c}$.

The interaction between two non-neutral and uniformly charged
surfaces is obtained by setting $\mathbf{k}_{0}=\varphi=0$ in
\eqs~(\ref{eq:decoupled_free_energy})-(\ref{eq:single_mode_sigma_2}), recovering the well-known result of Parsegian and Gingell
\cite{gingell1972}:
\begin{equation}
\frac{ F\left(d\right)}{A}=
\frac{\mathrm{e}^{-\kd d/2}}{\sinh \left(\kd d/2\right)}C_{+}^{2}
-\frac{\mathrm{e}^{-\kd d/2}}{\cosh \left(\kd d/2\right)}C_{-}^{2}
\,,\label{eq:PG_energy}
\end{equation}
written here in a decoupled way, separating the pure repulsive
from the pure attractive contributions in terms of the amplitudes
$C_{\pm}=\left(C_{\sigma}\pm C_{\eta}\right)/2$, where $C_{\sigma}$
and $C_{\eta}$ are uniform surface charge densities.
Unlike neutral modulated surfaces, for two uniformly charged surfaces
with the same sign ($C_\sigma\cdot C_\eta>0$),
the interaction is purely repulsive since the inequality $C_{+}>C_{-}$ is always valid.

For uniform charged surfaces, only
the ions in solution screen the interaction, whereas for modulated charged surfaces
there is an additional screening effect, yielding an effective screening length, $q_{0}^{-1}=1/\sqrt{\kd^{2}+k_0^2}$. Therefore, the lateral charge modulations result in a faster decay of the interaction as compared to the interaction between two uniformly charged surfaces, due to
the additional mechanism of screening.

Note that when two surfaces have each a different single-mode modulation, ${k}_\sigma \neq {k}_\eta$,
the integral over the coupling terms in \eq~(\ref{eq:dim_free_energy_2}) vanishes, leading to a purely repulsive interaction:
\begin{equation}
\frac{ F\left(d\right)}{A}=
\frac{\kd}{4 q_{\sigma}}\frac{\mathrm{e}^{-q_{\sigma}d}}{\sinh(q_{\sigma}d)}C_{\sigma}^{2}
+\frac{\kd}{4 q_{\eta}}\frac{ \mathrm{e}^{-q_{\eta}d}}{\sinh(q_{\eta}d)}C_{\eta}^{2}\,,
\label{eq:diff_k_energy}
\end{equation}
where $q_\sigma=\sqrt{\kd^2+k_\sigma^2}$, $q_\eta=\sqrt{\kd^2+k_\eta^2}$, and ${k}_{\sigma}$, ${k}_{\eta}$ are the wavenumbers of the
bottom and top surfaces, respectively.

The inter-surface interaction as expressed in
\eq~(\ref{eq:diff_k_energy}) can be interpreted as
a superposition of two decoupled systems each composed of
one uniform charged surface and a second (virtual) surface that is neutral. Namely, the first term corresponds
to  a uniformly charged surface  located at $z=-d/2$
and a neutral one located at $z=+d/2$, while the second term corresponds
to  a neutral surface located at $z=-d/2$
and a uniformly charged surface located at $z=+d/2$.
Although the \emph{virtual} surface
is neutral and does not interact electrostatically, it leads to a repulsive interaction
because the counter-ions are confined between the two surfaces and  their translational entropy is reduced.
The magnitude of the confined ionic volume depends linearly on $d$, and
decreases at smaller separations.

The above result suggests that a slight difference between the two wavelengths (${k}_{\sigma}\neq {k}_{\eta}$)
is sufficient to eliminate any inter-surface attraction between two periodically ordered surfaces,
implying that the attraction is a delicate effect that might be difficult
to obtain in experimental conditions corresponding to the linear PB regime.

\subsection{Multi-mode Modulation \label{sub:Neutral-surface-with}}

The analysis of the previous section can be generalized to multi-mode
distributions of the charges on the two surfaces. The charge distributions are written as:
\begin{eqnarray}
 \sigma\left(\boldsymbol{\rho}\right)&=& \delta\sigma\left(\boldsymbol{\rho}\right)+ \sigma_{0}\,\,\nonumber\\
 \eta\left(\boldsymbol{\rho}\right) & = & \delta\eta\left(\boldsymbol{\rho}\right) + \eta_{0}\label{eq:fluc_charge}
\,,
\end{eqnarray}
where $\sigma_{0}=\langle\sigma\rangle$ and $\eta_{0}=\langle\eta\rangle$ are the average surface charge
densities, and $\delta\sigma\left(\boldsymbol{\rho}\right)$ and $\delta\eta\left(\boldsymbol{\rho}\right)$
are the charge modulations around these averages.
The inter-surface interaction of Eq.~(\ref{eq:dim_free_energy_2}) can be separated into two terms
\begin{align}
 F\left(d\right)= &  F_{0}\left(d\right)+ F_{1}\left(d\right)\,,
\end{align}
where the first term depends only on $\sigma_0$ and $\eta_0$, and coincides with Eq.~(\ref{eq:PG_energy}) with $C_\pm=(\sigma_0\pm\eta_0)/2$:
\begin{equation}\label{eq:unifrom_energy}
\frac{ F_{0}\left(d\right)}{A}=
\frac{2\sigma_{0}\eta_{0}+\left(\sigma_{0}^{2}+\eta_{0}^{2}\right)\mathrm{e}^{-\kd d}}{2\sinh\left(\kd d\right)}
\, .
\end{equation}
%
It vanishes when both surfaces are neutral, $\sigma_0=\eta_0=0$. The second term
is the contribution due to charge heterogeneity, given by Eq.~(\ref{eq:dim_free_energy_2}) without the mode $\mathbf{k}=0$:
\begin{align}\label{eq:general_F_1}
 F_{1}\left(d\right)=
\kd\int_{\mathbf{k}\neq0}\frac{\mathrm{d}^{2}k}{\left(2\pi\right)^{2}}
\left[\frac{{\delta\sigma}_{\mathbf{k}}{\delta\eta}_{-\mathbf{k}}
+{\delta\sigma}_{-\mathbf{k}}{\delta\eta}_{\mathbf{k}}}
{2q\sinh\left(qd\right)}\right.
\quad\quad\quad
\nonumber
\\
+\left.\frac{\left({\delta\sigma}_{\mathbf{k}}{\delta\sigma}_{-\mathbf{k}}
+{\delta\eta}_{\mathbf{k}}{\delta\eta}_{-\mathbf{k}}\right)
\mathrm{e}^{-qd}}{2q\sinh\left(qd\right)}\right]~,
\end{align}
where ${\delta\sigma}_{\mathbf{k}}$ and ${\delta\eta}_{\mathbf{k}}$ are the Fourier transforms of
$\delta\sigma\left(\boldsymbol{\rho}\right)$ and $\delta\eta\left(\boldsymbol{\rho}\right)$, respectively.
By definition, from Eq.~(\ref{eq:fluc_charge}), $\langle\delta\sigma\rangle=\langle\delta\eta\rangle=0$,
or equivalently in Fourier space ${\delta\sigma}_{\mathbf{k}=0}={\delta\eta}_{\mathbf{k}=0}=0$. Note that
${\delta\sigma}_{\mathbf{k}}$ and ${\delta\eta}_{\mathbf{k}}$ have units of length in Fourier space (unlike the units of
$C_\sigma$ and $C_\eta$). Hence, $F_1$ is dimensionless.

An important length scale for a general periodic (multi-mode) distribution is periodic charge domain size, $L$,
which is related to the smallest wavenumber, $L=2\pi/{k}_{\mathrm{min}}$.
The wavevector $\mathbf{k}_\mathrm{min}$ also
acts as a lower cutoff in the $\mathbf{k}$-space integration, \eq~(\ref{eq:general_F_1}).

Two limiting cases of small and large  ${k}_\mathrm{min}$ can be considered separately. In
the limit of charges with molecular-size heterogeneities, the domain size is molecular $L\ll \kd^{-1}$,
and $q_{\mathrm{min}}=\sqrt{\kd^2+k_\mathrm{min}^2}\simeq2\pi/ L\gg\kd$ depends inversely on $L$.
Therefore, the leading term in the integrand varies as $ \kd q^{-1}\mathrm{e}^{-qd}\simeq\kd L\e^{-2\pi d/L}\rightarrow 0$
(as long as $L\ll d$), and leads to a negligible contribution due to charge heterogeneity,
$F_{1}\left( d\right)\rightarrow0$.

On the
other hand, in the limit of large domains, $L\gg \kd^{-1}$,
$q_\mathrm{min}\simeq \kd+O\left(L^{-2}\right)$ has no dependence
on $L$. The decay length of the interaction is the Debye length, $\kd^{-1}$,
similarly to the average charge contribution in \eq~(\ref{eq:unifrom_energy}).
The leading term of $F_1(d)$ is then given by:
\begin{equation}\label{eq:large_domain_energy}
 F_{1}\left(d\right)\simeq\frac{S_{1}+S_{2}\mathrm{e}^{-\kd d}}{\sinh\left(\kd d\right)}\,,
\end{equation}
where the prefactors\\
$S_{1}=\frac{1}{2}
\int\frac{\mathrm{d}^2 k}{(2\pi)^2}(\delta\sigma_{\mathbf{k}}\delta\eta_{-\mathbf{k}}
+\delta\sigma_{-\mathbf{k}}\delta\eta_{\mathbf{k}})$\,,
and
$S_{2}=\frac{1}{2}\int\frac{\mathrm{d}^2 k}{(2\pi)^2}(\left|\delta\sigma_\mathbf{k}\right|^2
+\left|\delta\eta_\mathbf{k}\right|^2)>0$\,
do not depend on $d$, and can be interpreted as contributions from effective uniform surface
charges.

The similar dependence on $d$ of Eqs.~(\ref{eq:unifrom_energy}) and (\ref{eq:large_domain_energy}) implies that the effect of large domains can be understood as a modification of the prefactors in \eq~(\ref{eq:unifrom_energy}).
When $S_1$ is negative, the contribution $ F_{1}\left(d\right)\simeq 2 S_1\e ^{-\kd d}<0$,
is attractive for large separations $\kd d\gg 1$,
and is followed by a crossover at $d_c=\kd^{-1}\cosh^{-1}\left(-S_2/S_1\right)$
from attraction for $d>d_c$ to repulsion for $d<d_c$, resembling the crossover found in Eq.~(\ref{eq:single_mode_energy_2}).

Examining the two limits of small and large domains leads to the observation that periodic charge modulation has a significant effect only when the domain size is sufficiently large. Thus, for charge modulations
 characterized by molecular heterogeneities, one can treat the interaction effectively as if the charges on the two surfaces are distributed homogeneously. On the other hand, for surfaces where the charge domains are larger than the Debye screening length $\kd^{-1}$, the contribution due to charge modulation cannot be neglected, and may lead to a substantial change in the strength of the inter-surface interaction. Furthermore, a crossover from repulsion
 at $d<d_c$ to attraction at larger separations, $d>d_c$, can be induced.

 \subsection{Time-dependent Lateral Displacement and Inter-surface Correlations}

The case of two identical single-mode distributions with a relative phase $\varphi$, \eqs~(\ref{eq:single_mode_sigma_1})-(\ref{eq:single_mode_energy_2}),
can be generalized for two identical multi-mode distributions with a relative lateral displacement
$\boldsymbol{\ell}$:
\begin{equation}\label{eq:identical_dispalced_charges}
\delta\eta\left(\boldsymbol{\rho}+\boldsymbol{\ell}\right)=\delta\sigma\left(\boldsymbol{\rho}\right)\,.
\end{equation}
The above charge distribution, Eq.~(\ref{eq:identical_dispalced_charges}), is motivated by a specific inter-surface force experiment setup~\cite{silbert2011} that will be discussed in Sec.~V, while here we explore its theoretical consequences.
The relative displacement $\boldsymbol{\ell}=(\ell_x,\ell_y)$ is related to a ${k}$-dependent phase $\varphi_{\mathbf{k}}=\mathbf{k}\cdot\boldsymbol{\ell}$ in Fourier space with $\delta\eta{}_{\mathbf{k}}=\delta\sigma{}_{\mathbf{k}}\mathrm{e}^{i\varphi_\mathbf{k}}$, and $ F_{1}$ from \eq~(\ref{eq:general_F_1}) reads:
\begin{equation}
 F_{1}\left(d\right)=
\kd\int\frac{\mathrm{d}^{2}k}{\left(2\pi\right)^{2}}
\frac{\cos\left(\mathbf{k}\cdot\boldsymbol{\mathbf{\ell}}\right)
+\mathrm{e}^{-qd}}{q\sinh\left(qd\right)}\left|\delta\sigma_{\mathbf{k}}\right|^{2}
\,.\label{eq:modul_free_energy}
\end{equation}
For $\boldsymbol{\ell}=0$ the two distributions are in-phase (symmetric), leading to repulsive contributions from all modes for any $d$. In the more general case of a relative displacement $\boldsymbol{\ell}\ne 0$, each $\mathbf{k}$-mode can have either an attractive or a repulsive contribution, and \eq~(\ref{eq:modul_free_energy}) gives the overall positive/negative sign of $F_1$, due to strong correlations between the two surfaces.

We consider two surfaces with identical charge distributions that are relatively displaced by applying on one of them a time-dependent lateral force. This force leads to a time-dependent lateral vector $\boldsymbol{\ell}(t)$
such that $\delta\eta(\boldsymbol{\rho}+\boldsymbol{\ell}(t))=\delta\sigma(\boldsymbol{\rho})$.
The time dependence $\boldsymbol{\ell}(t)$ is related to the temporal dependence of the applied lateral (shear) force. The validity of this model holds as long as the lateral motion is a quasi-static process, for which the ions in the solution equilibrate faster than the typical time-scale of the surface lateral motion.
Then, the inter-surface interaction is given by averaging the
$\cos\left[\mathbf{k}\cdot \boldsymbol{\ell}(t)\right]$ term in \eq~(\ref{eq:modul_free_energy}) over the
displacement period $T$ :
\begin{equation}\label{eq:time_dep_energy}
\left\langle  F_{1}\left(d\right)\right\rangle _{T}=\kd\int\frac{\mathrm{d}^{2}k}
{\left(2\pi\right)^{2}}\frac{\left\langle \cos\left[\mathbf{k}\cdot\boldsymbol{\ell}(t)\right]\right\rangle _{T}+\mathrm{e}^{-qd}}{q\sinh\left(qd\right)}\left|\delta\sigma_{\mathbf{k}}\right|^{2}
\,,
\end{equation}
where $F_1$ is the contribution to the inter-surface interaction due to charge modulation, and
$\left\langle O\right\rangle _{T}=T^{-1}\int_0^T\mathrm{d}t\,O(t)$ is the time average.
Since the average of the cosine depends on the lateral periodic motion, the
interaction energy would vary when changing the oscillatory mode of the force.

A simple example is a square-wave motion in the $\hat{x}$ direction, given by:
\begin{equation}
\label{eq:sq_wave}
\boldsymbol{\ell}(0<t<T)=
\begin{cases}
\left(\ell_0- {\Delta \ell}\right)\hat{x} &~~~ 0<t\le T/2,
\\
\left(\ell_0+{\Delta \ell}\right)\hat{x} &~~~ T/2<t \le T,
\end{cases}
\end{equation}
where ${\ell}_0$ is the mean lateral displacement between the two surfaces,
and $\Delta {\ell}$ is the oscillation amplitude.
The inter-surface interaction is given by
\begin{align}
\left\langle  F_{1}\left(d\right)\right\rangle _{T}=
\quad\quad\quad\quad\quad\quad\quad\quad\quad\quad\quad\quad\quad\quad\quad\quad\quad\quad
\nonumber
\\
\kd\int\frac{\mathrm{d}^{2}k}{\left(2\pi\right)^{2}}\frac{ \cos\left({k}_x {\ell}_0\right)\cos\left({k}_x {\Delta\ell}\right) +\mathrm{e}^{-qd}}{q\sinh\left(qd\right)}\left|\delta\sigma_{\mathbf{k}}\right|^{2}
\, ,
\label{eq:lateral_1}
\end{align}
and depends both on the mean displacement $\ell_0$ and the oscillation amplitude $\Delta\ell$.
Here $k_x$ is the wavevector component along the direction of the shear motion.

For a general periodic dependence ${\ell}(t)$, if the amplitude of the lateral oscillations $\Delta\ell(t)=\ell(t)-\ell_0$ is small, ${k}_x\Delta{\ell} \ll 1$, and to leading orders
the interaction depends on the average displacement, ${\ell}_{0}$,
and the mean square amplitude of the oscillation, $\langle\Delta\ell^2\rangle_{T}=\Delta \ell^2$:
\begin{align}
\left\langle  F_{1}\left(d\right)\right\rangle_{T}\simeq
\kd\int\frac{\mathrm{d}^{2}k}{\left(2\pi\right)^{2}}
\left[
\frac{(1-\frac{1}{2} k_x^2 (\Delta\ell)^2)\cos(k_x\ell_0)}{q\sinh(qd)}\right.
\nonumber
\\
+~~\frac{\mathrm{e}^{-qd}}{q\sinh\left(qd\right)}\Bigg{]}\left|\delta\sigma_{\mathbf{k}}\right|^{2}
\,.\nonumber\\
\label{eq:lateral_2}
\end{align}
Note that the above equation is exact for the square-wave distribution of \eq~(\ref{eq:sq_wave}).

\begin{figure*}
\begin{centering}
\includegraphics[width=0.35\textwidth]{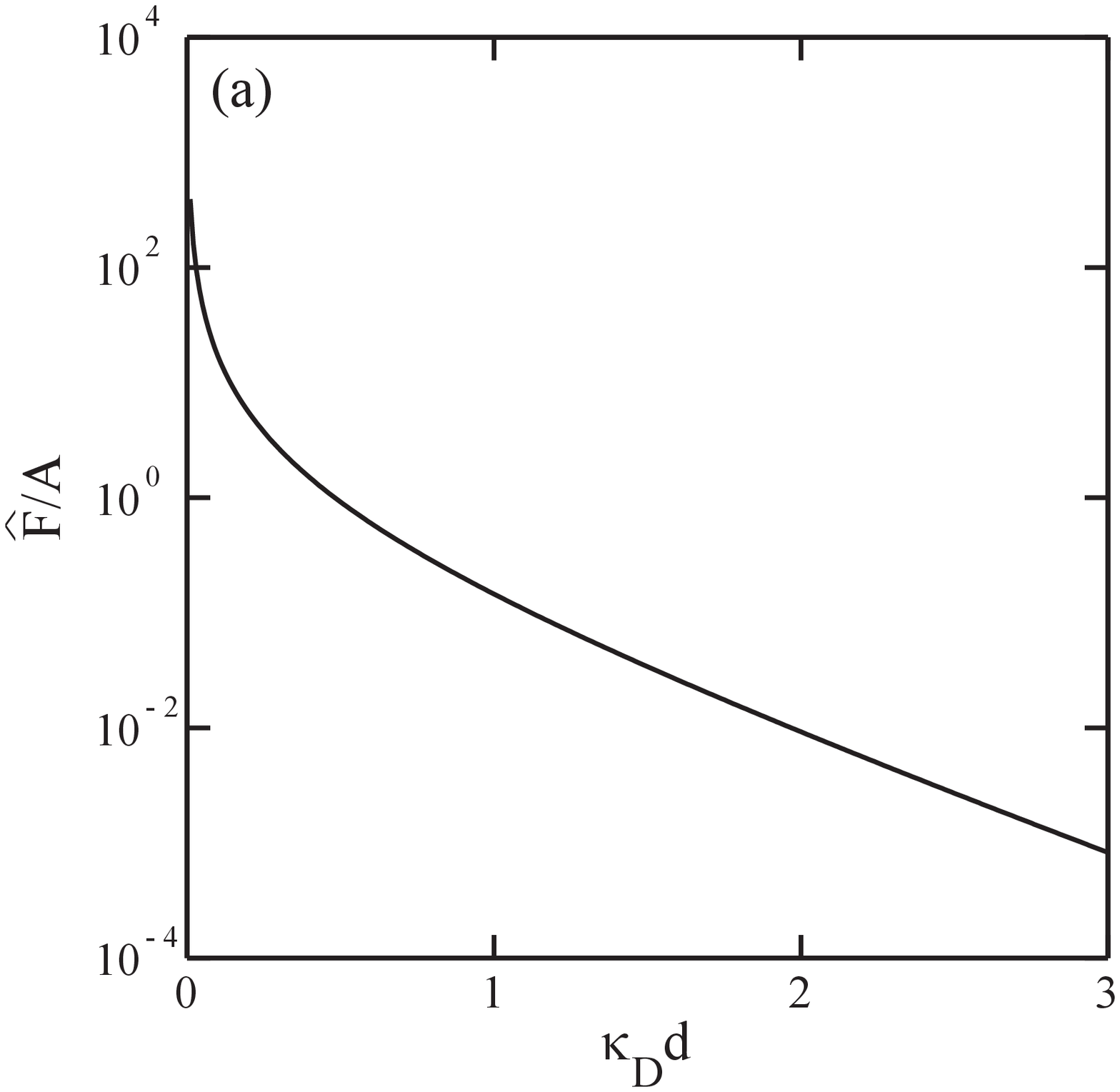} ~~~~~~~ \includegraphics[scale=0.38]{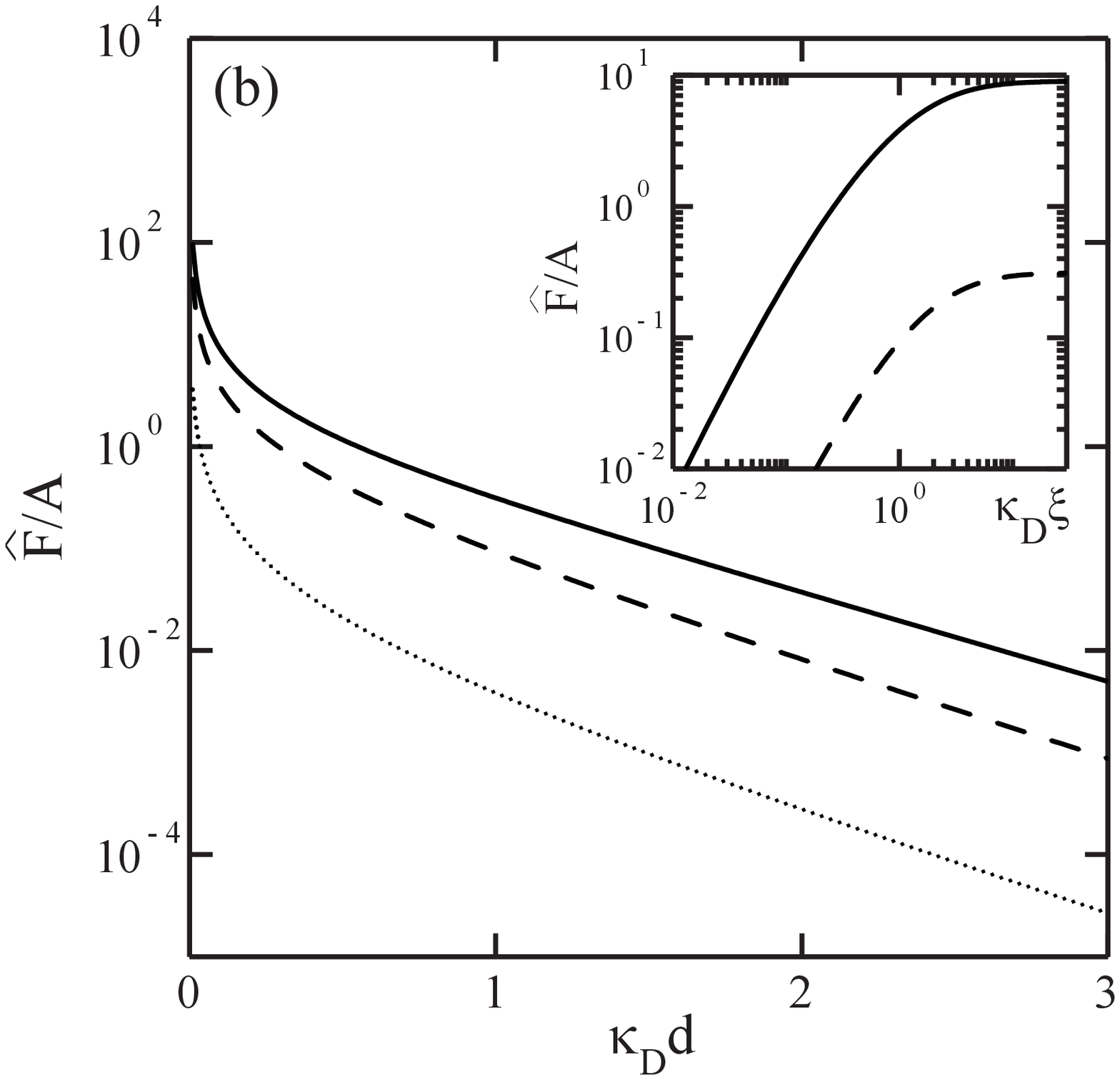}
\par\end{centering}
\caption{\label{fig:I(d)}The averaged interaction energy, $\widehat{F}/A$ as a function of the inter-surface separation $d$
for randomly charged surfaces. In (a)
the free energy from \eq~(\ref{eq:uncorrelated_interaction}), for an
uncorrelated Gaussian distribution \eq~(\ref{eq:gaus_corr}),
is plotted on a semi-log plot as a function of $\kd d$.
In (b) the free energy is plotted on a semi-log plot as a function of $\kd d$, where the charge
distribution is a two-dimensional Lorentzian, \eq~(\ref{eq:lorentzian}).
Three values of $\kd \xi$ are shown: $\kd\xi=100$ (solid line), $\kd\xi=1$ (dashed line), and $\kd\xi=0.1$ (dotted line).
In the inset the dependence of the free energy on $\kd\xi$
is plotted on a log-log scale for two values of the separation, $\kd d=0.1$ (solid line),
and $\kd d=1$ (dashed line). The plotted free energy is rescaled by the prefactor of \eq~(\ref{eq:uncorrelated_interaction}) $\widehat{F}=2\pi \left\langle  F\right\rangle /\left(\kd^2 \gamma^2 a^2\right)$ for (a), and $\widehat{F}=\left\langle  F\right\rangle /\gamma^2$ from Eq.~(\ref{eq:lorentzian}) for (b).}
\end{figure*}

\section{Quenched Surface Disorder: Patchy Surfaces\label{sub:Random-quenched-disorder}}
We generalize now the periodic distribution of Sec.~III  to patchy surfaces with random charge domains, but which are overall neutral.
Consider two surfaces with charge distributions $\sigma\left(\boldsymbol{\rho}\right)$
and $\eta\left(\boldsymbol{\rho}\right)$, respectively.  The surface charges are randomly distributed with
a joint probability distribution $P\left[\sigma\left(\boldsymbol{\rho}\right),\eta\left(\boldsymbol{\rho}\right)\right]$.
The inter-surface interaction, \eq~(\ref{eq:dim_free_energy_2}), is obtained
by averaging over the bilinear terms of the Fourier transform:
\begin{align}
\left\langle  F\left(d\right)\right\rangle _{\sigma,\eta} & =
\kd\int\frac{\mathrm{d}^{2}k}{\left(2\pi\right)^{2}}\,
\left[\frac{\left\langle {\sigma}_{\mathbf{k}}{\eta}_{-\mathbf{k}}\right\rangle _{\sigma,\eta}+\left\langle {\sigma}_{-\mathbf{k}}{\eta}_{\mathbf{k}}\right\rangle _{\sigma,\eta}}{2q\sinh\left(qd\right)}\right.
\quad\quad\quad
\nonumber
\\
&
\quad\quad+\left.\frac{\left\langle {\sigma}_{\mathbf{k}}{\sigma}_{-\mathbf{k}}\right\rangle _{\sigma,\eta}+\left\langle {\eta}_{\mathbf{k}}{\eta}_{-\mathbf{k}}\right\rangle _{\sigma,\eta}}{2q\sinh\left(qd\right)}\mathrm{e}^{-qd}\right] \,,
\end{align}
where $\left\langle O\right\rangle _{\sigma,\eta}=\int{\cal D} \sigma \int{\cal D}
\eta P\left[\sigma,\eta\right]O$ is the average over the joint $\sigma$ and $\eta$ distribution.
In experiments the probability distribution is usually determined
by the preparation procedure of the surfaces.

When the surfaces are
prepared separately, there are no inter-surface correlations,
$\left\langle {\sigma}_{\mathbf{k}}{\eta}_{-\mathbf{k}}\right\rangle =0$,
leading to a purely repulsive interaction:
\begin{align}
\left\langle  F\left(d\right)\right\rangle _{\sigma,\eta}=
\quad\quad\quad\quad\quad\quad\quad\quad\quad\quad\quad\quad\quad\quad
\nonumber
\\
\kd\int\frac{\mathrm{d}^{2}k}{\left(2\pi\right)^{2}}\,\frac{\left\langle {\sigma}_{\mathbf{k}}{\sigma}_{-\mathbf{k}}\right\rangle _{\sigma}+\left\langle {\eta}_{\mathbf{k}}{\eta}_{-\mathbf{k}}\right\rangle _{\eta}}{2q\sinh\left(qd\right)}\mathrm{e}^{-qd}\,>\,0\,,
\end{align}
and the strength of the repulsive interaction depends on the probability
distribution of each surface. For an uncorrelated Gaussian
distribution, the two-point correlation function is given by:
\begin{equation}\label{eq:gaus_corr}
\left\langle \sigma\left(\boldsymbol{\rho}\right)\sigma\left(\boldsymbol{\rho}'\right)\right\rangle _{\sigma}=\left\langle \eta\left(\boldsymbol{\rho}\right)\eta\left(\boldsymbol{\rho}'\right)\right\rangle _{\eta}=\gamma^{2}\delta\left(\frac{\boldsymbol{\rho}-\boldsymbol{\rho}'}{a}\right)\,,
\end{equation}
where $\delta(\boldsymbol{\rho})$ is the two-dimensional Dirac $\delta$-function, $\gamma$ is the root mean square charge density (taken to be the same on the two surfaces), and $a$
is a conveniently defined molecular length. This leads to $\left\langle {\sigma}_{\mathbf{k}}{\sigma}_{-\mathbf{k}}\right\rangle _{\sigma}=\left\langle {\eta}_{\mathbf{k}}{\eta}_{-\mathbf{k}}\right\rangle _{\eta}= A\gamma^{2}a^{2}$, and
to an inter-surface interaction per unit area:
\begin{equation}
\frac{\left\langle  F \left(d\right)\right\rangle _{\sigma,\eta}}{A}=\frac{\kd^2 \gamma^{2}a^{2}}{2\pi}I\left(\kd d\right) >0
\,,\label{eq:uncorrelated_interaction}
\end{equation}
where
\begin{equation}
I\left(\kd d\right)=\int_{0}^{\infty}\mathrm{d}k\,\frac{k\mathrm{e}^{-qd}}{\kd q\sinh\left(qd\right)}
=-\frac{\ln\left(1-\e^{-2\kd d}\right)}{\kd d}
\,.\label{eq:gaussian_integral}
\end{equation}
The repulsive free-energy for the uncorrelated Gaussian case is presented in \fig~\ref{fig:I(d)}(a).
At separations larger than the Debye screening length, $\kd d\gg 1$,  the
leading term decays exponentially as $\left\langle  F \right\rangle{/}A\sim\mathrm{e}^{-2\kd d}/\left(\kd d\right)$, while at
small separations, $\kd d\ll 1$, the energy scales as $\left\langle  F \right\rangle{/}A\sim\ln\left[1/(2\kd d)\right]/\left(\kd d\right)$.

When calculating these scaling relations we ignored the molecular length scale $a$, since $I(\kd d)$ has no dependence on $a$. However, the prefactor in \eq~(\ref{eq:uncorrelated_interaction}) scales as the square of the molecular length scale ${  \left\langle  F \right\rangle }/{A}\sim \left(\kd a\right)^{2}$. Hence, the contribution of uncorrelated Gaussian disorder, \eq~(\ref{eq:gaus_corr}), vanishes in the limit where $a$ is much smaller than the Debye length,
$\kd a\ll 1$, in agreement with the results reported in Ref.~\cite{naji2005}.

In order to model the possibility of finite (macroscopic) charge domains we replace \eq~(\ref{eq:gaus_corr}) by a two-point correlation function having a Lorentzian
distribution:
\begin{equation}
\left\langle \sigma\left(\boldsymbol{\rho}\right)
\sigma\left(\boldsymbol{\rho}'\right)\right\rangle _{\sigma}=\left\langle \eta\left(\boldsymbol{\rho}\right)\eta\left(\boldsymbol{\rho}'\right)\right\rangle _{\eta}=\frac{\gamma^{2}\xi^{2}}{\left(\boldsymbol{\rho}-\boldsymbol{\rho}'\right)^{2}+\xi^{2}}\,,
\end{equation}
where $\xi$ is the charge correlation length on the surface, which can be associated with the domain size.
In Fourier space this leads to
$\left\langle {\sigma}_{\mathbf{k}}{\sigma}_{-\mathbf{k}}\right\rangle _{\sigma}=\left\langle {\eta}_{\mathbf{k}}{\eta}_{-\mathbf{k}}\right\rangle _{\eta}=2\pi A\gamma^{2}\xi^{2}K_{0}\left(k\xi\right)$,
where $K_{0}\left(x\right)$ is the zeroth-order modified Bessel function of the
second kind.

The inter-surface interaction  is then given by:
\begin{equation}
\frac{\left\langle  F \left(d\right)\right\rangle _{\sigma,\eta}}{A}=
\kd\gamma^{2}\xi^{2}\int_{0}^{\infty}\mathrm{d}k\,\frac{k\mathrm{e}^{-qd}K_{0}\left(k\xi\right)}{q\sinh\left(qd\right)}
\,,\label{eq:lorentzian}
\end{equation}
and the dependence of $\langle F(d)\rangle_{\sigma,\eta}$ on $\kd d$ for three values of
the correlation length $\xi$ is shown in \fig~\ref{fig:I(d)}(b). In the
limit of short-range correlations [dotted line in \ref{fig:I(d)}(b)], $\kd\xi\ll 1$, the integrand in \eq~(\ref{eq:lorentzian})
depends weakly on $\xi$, and the interaction energy scales as
$\left\langle  F \right\rangle/{A}\sim-\ln(\kd \xi)\xi^{2}$, resembling the scaling of the Gaussian distribution case, \eq~(\ref{eq:uncorrelated_interaction}), where $\left\langle  F \right\rangle\sim a^2$.
In the other limit of large domains [solid line in \ref{fig:I(d)}(b)], where $\kd\xi\gg 1$,
the main contribution to the integral comes from the small values of $k$,
and the interaction is independent of $\xi$, $\left\langle  F \right\rangle/{A}\sim\e^{-\kd d}/\sinh\left(\kd d\right)$. {This reproduces repulsion with a decay length that does not depend on the lateral length scale, $\xi$. It is effectively analogous to repulsion between uniform surfaces as in \eq~(\ref{eq:PG_energy}) with an effective separation that is twice as large as that of the homogeneous case.} These two limiting behaviors
of $ \left\langle  F \right\rangle/A$ as a function of $\kd \xi$ are shown in the inset of \fig~\ref{fig:I(d)}(b) for two values of $d$.

By examining the short- and long-range charge distributions we find that short-range charge disorder has a negligible effect on the inter-surface energy, while long-range disorder leads to a substantial effect, which cannot be omitted even in the case of non-zero average charge. This observation is in accord with the results presented in the multi-mode periodic case, Sec.~\ref{sub:Neutral-surface-with}, where it was found that charge distributions with large charge domains changed the inter-surface interaction substantially, while distributions with small domains lead to a negligible effect.

\section{Discussion}\label{sec:discussion}

The model we present accounts for the inter-surface interaction of two
heterogeneously charged surfaces. It is formulated on the mean-field
level where the electrostatic potential is given by the Poisson-Boltzmann (PB) equation.
Our main aim is to study the effect of quenched charge
disorder with finite domains on the inter-surface interaction between the two surfaces.

We focus on the linear regime of the PB equation, where it is found that the inter-surface interaction  of two arbitrary charge distributions depends on a bilinear coupling between the Fourier components of the two distributions. It is shown that the interaction can be expressed as a sum of an attractive term and a repulsive one.
The repulsive term at small separations is usually larger than the attractive one, {leading to a repulsion for $d\rightarrow 0$.} In a special case where the two distributions are exactly out-of-phase, $\sigma\left(\boldsymbol{\rho}\right)=-\eta \left(\boldsymbol{\rho}\right)$, such that negatively charge domains are located against positively charge domains in the other surface and vice versa, the interaction is attractive for any value of $d$.

Several types of surface charge distributions are studied in the linear PB regime.
First, we consider two overall neutral surfaces with the same single-mode periodic charge density, but
with a relative phase shift $\varphi$ between the two.
When the average surface charge is taken to be zero,
the inter-surface interaction stems only from the lateral charge
modulation.
When the top surface mode is identical with the bottom one,
the interaction can vary between pure attraction and pure
repulsion as function of the relative phase $\varphi$ between the two surfaces.
For an out-of-phase distribution, $\varphi=\pi$, there is a relative displacement
by half  a wavelength between the bottom and top surfaces, and the interaction
is purely attractive, while for the in-phase case, $\varphi=0$, it is purely
repulsive.

These two limits are a consequence of strong inter-surface correlations.
For intermediate values of the relative
phase $\varphi$, the correlation is weaker and a crossover from attraction (at large separations) to
repulsion (at small separations) occurs at a separation $d_{c}\sim\cosh^{-1}\left(1/\left|\cos\varphi\right|\right)$.
Note that the crossover separation $d_c$ does not depend on the modulation amplitude.
In the more general case where the bottom and top surfaces have different (single)
$k$-mode, the interaction is found to be purely repulsive.
The origin of the repulsion
in this case is due to the confinement of the counter-ions in between the surfaces.

The role of the domain size in a general periodic (multi-mode) distribution is also investigated. It is found that the contribution due to periodic charge modulation is important only in the limit of large domains $\kd L\gg 1$. For sufficiently large domains the contribution due to
charge heterogeneity is substantial, even in the case of non-neutral surfaces, where the main contribution stems from the average surface charge.

The time-dependent relative displacement case discussed in Sec.~III.C is motivated by recent surface force experiment \cite{silbert2011}, which examined whether the charge heterogeneity on the two surfaces is annealed or randomly distributed (quenched case).
In the experiments, the normal forces between two mica surfaces partially coated with a cationic lipid bilayer were measured, while during the vertical approach the surfaces were also sheared laterally in an oscillatory mode at a rate which is slower than the typical time required for the ions in solution to rearrange, but faster than the time scale of vertical approach or lipids rearrangement on the coated surfaces. Thus, it is assumed that the initial ordering of the surface charge on each surface is preserved during the normal approach.

If the hypothesis that the surfaces are periodically ordered \cite{brewster2008} is correct, then the measured inter-surface force should depend on the average and mean square amplitude of the lateral motion as in \eqs~(\ref{eq:lateral_1})-(\ref{eq:lateral_2}). However, it was shown in Ref.~\cite{silbert2011} that the forces measured for different lateral motions remained unchanged, suggesting that the surface charge distribution is not periodic, but presumably is quenched and random.


While in the annealed case the surface charges can rearrange themselves,
and the system approaches the minimum energy state as was investigated
in Refs.~\cite{naydenov2007,brewster2008,jho2011,deLaCruz2006},
in the quenched case the accessible configurations of the surfaces are frozen.
The quenched probability distribution is usually determined by
the experimental setup. For example, in surface force experiments the two surfaces are prepared separately before the force measurement, and their corresponding charge distributions can be assumed to be independent with no inter-surface correlation.
We consider this case and  find that the interaction is purely repulsive in the linear PB model,
due to the vanishing of inter-surface bilinear coupling terms.
{The inter-surface interaction can be thought of as an average over equal weights of repulsive and attractive contributions. However, since the repulsive term varies as $\exp(-qd/2)/\sinh(qd/2)$ and the attractive one as $\exp(-qd/2)/\cosh(qd/2)$, the repulsive contribution is always larger than the attractive one, leading to an overall repulsive interaction between uncorrelated randomly charged surfaces in the linear PB model.}

We discuss two specific types of random distributions. For Gaussian charge distributions with no inter- and intra-surface correlations, the interaction depends on a molecular length scale $a$ (lower cutoff length). Disorder at the molecular level may emerge due to local processes that are not affected by neighboring surface charges.
The contribution of the heterogeneity in this case is negligible, in agreement with Ref.~\cite{naji2005}.

In a more general case of Lorentzian distribution with correlation length $\xi$, we find two limiting regimes. For large values of $\xi$ we obtain a substantial effect due to charge disorder, while for small values of $\xi$ the limit of disorder at the molecular level is recovered and the contribution due to disorder is negligible.
{Note the similarity between these two limiting regimes and the case of periodic distributions mentioned earlier. For the latter and in the limit of large domains, $\kappa_{\rm D}L\gg 1$,  the effect is significant, while in the small domain limit, $\kappa_{\rm D}L\ll 1$, the domain-size effect is quite negligible.}

All the results obtained in the present work assumed a linear PB regime and demonstrate that quenched charge disorder leads (except in special set-ups) to a non-vanishing repulsive interaction between charged surfaces. It is reasonable to expect that similar considerations can be extended to the non-linear PB regime, where the electrostatic potential is too large for the linear approximation to be valid.
However, using a simplified model for the limit of infinitely-large charge domains, it was shown in Ref.~\cite{silbert2011} that the inter-surface interaction in the non-linear regime
may become overall attractive.

In the Appendix we reproduce this simplified treatment and compare its results, for a range of surface charge densities, with those of the linear model presented above.
For weakly charged surfaces, when the validity of the linear PB approximation can be justified, the overall repulsion predicted in the linear regime agrees with the non-linear one, while for strongly charged surfaces the linear PB equation fails to predict the attraction obtained in the non-linear regime.
This suggests a possible crossover from repulsive to attractive inter-surface interaction, as a function of charge density and patch size.

\section{Concluding Remarks}\label{sec:conclusion}

The results presented in our work demonstrate that charge heterogeneity may have important implications on the inter-surface interaction both for periodically modulated surfaces and randomly quenched ones. Several extensions to our work can be considered.

\begin{figure*}
\begin{centering}
\includegraphics[scale=0.4]{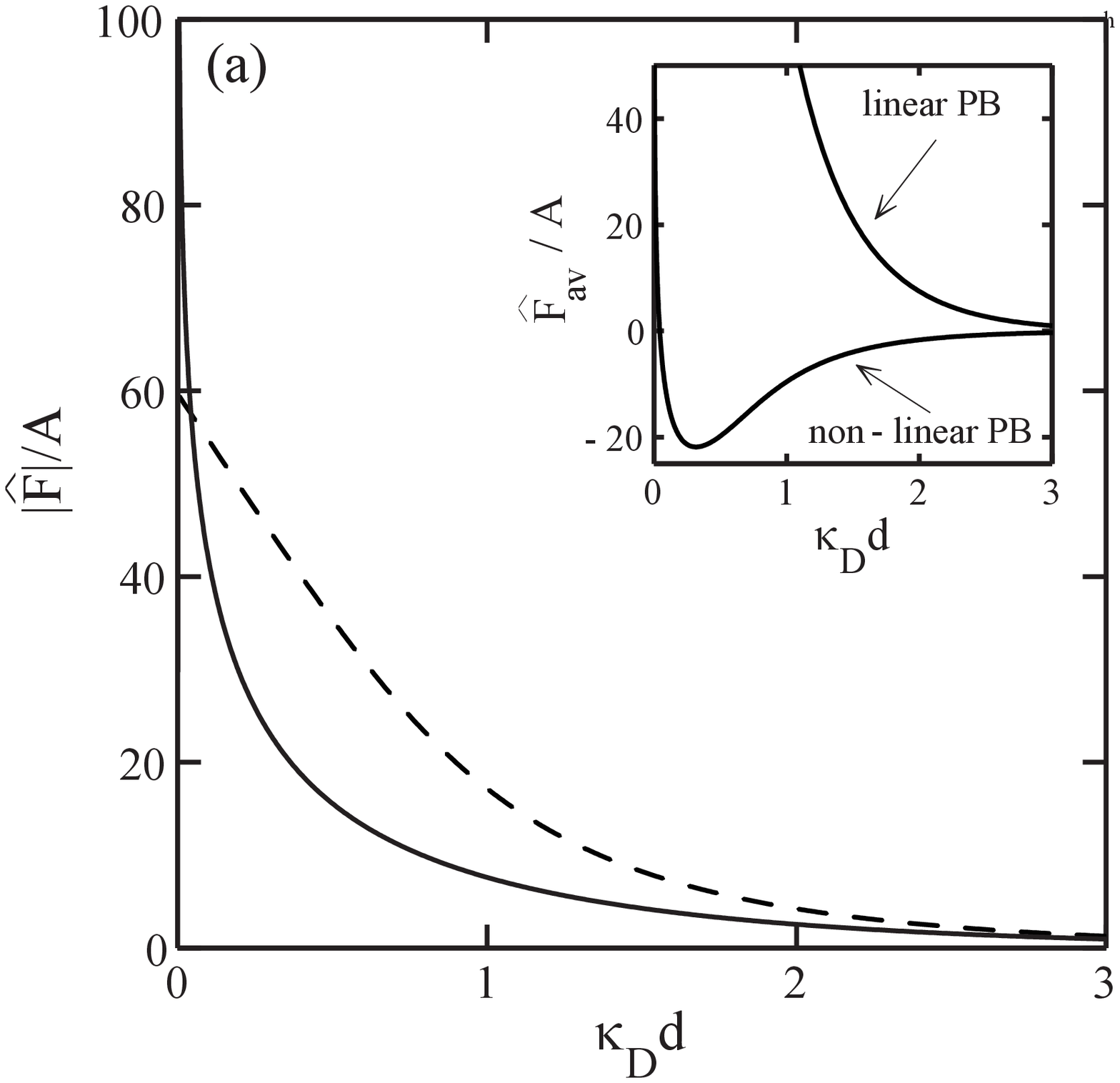} ~~~~~~~\includegraphics[scale=0.41]{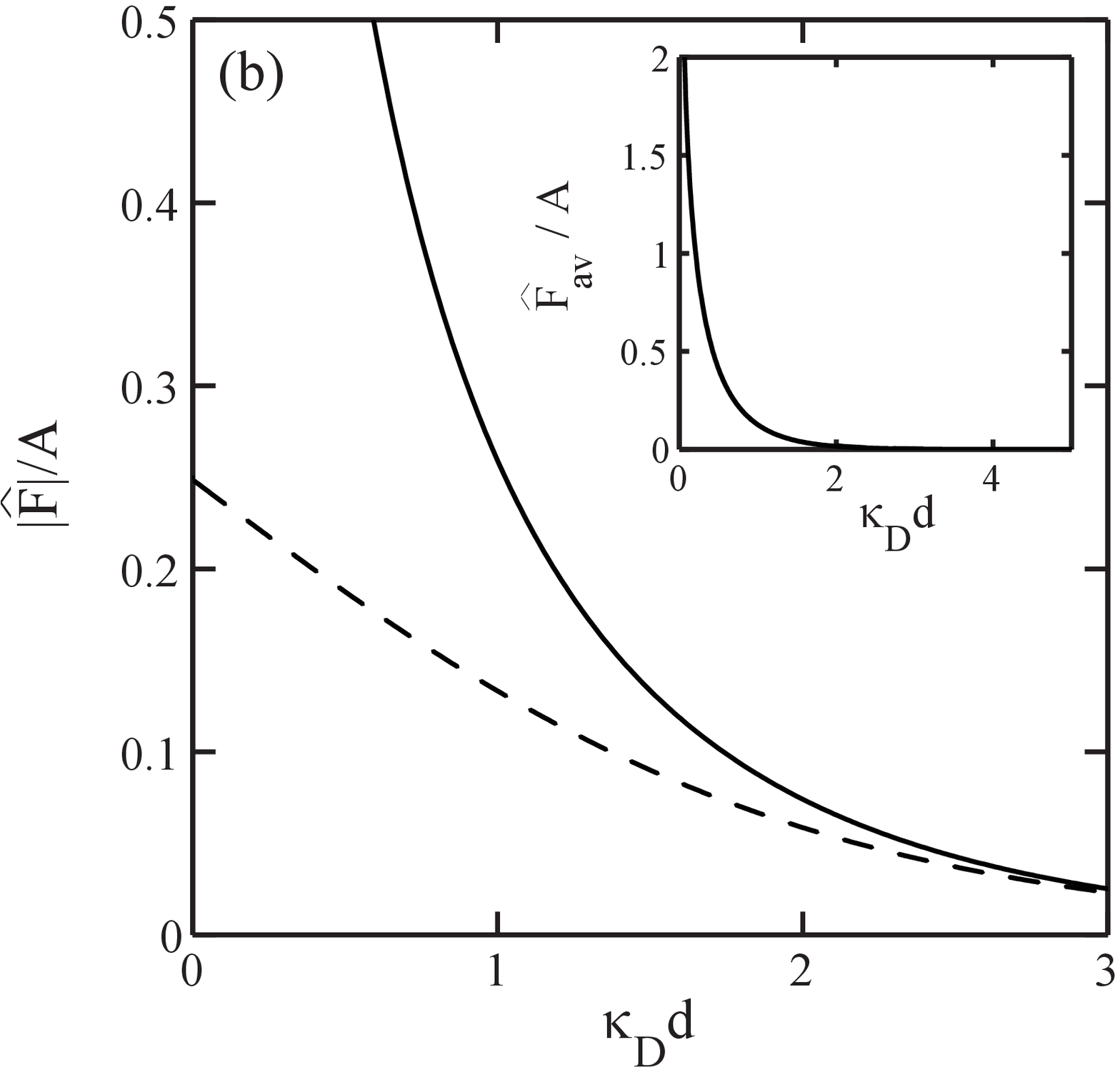}
\par\end{centering}
\caption{\label{fig:The-interaction-energy} The patch-patch inter-surface interaction energy per unit area, $|\widehat{F}|/A$, as a function of the separation $d$ rescaled by $\kd$. The plotted free energy $\widehat{F}=F/\kd^2$ is rescaled by $\kd^2$.
(a) Attraction $|\widehat{F}_\mathrm{att}|$ (dashed line) and repulsion $\widehat{F}_\mathrm{rep}$ (solid line) are calculated by the non-linear PB equation for $\sigma=-\eta=10\kd$ and $\sigma=\eta=10\kd$, respectively. The inset shows the average of the repulsive
and attractive corresponding free energies $\widehat{F}_{\rm av}=(\widehat{F}_\mathrm{rep}-|\widehat{F}_\mathrm{att}|)/2$, calculated by the linear and non-linear PB equation. (b) The solid and dashed lines are calculated with the non-linear PB equation for $\sigma=\eta=0.5\kd$,
and $\sigma=-\eta=0.5 \kd$, respectively. The inset shows the average $\widehat{F}_{\rm av}$. The results of the linear PB equation are omitted in (b) since their difference from the non-linear calculation is invisible.
}
\end{figure*}

{An effect not taken into account in the present work and left for future investigations is to look more carefully at the structure and thickness of the charged surface.
This may be of importance for charged membranes composed of mixtures of charged and neutral lipids. As the lateral membrane charge density may be correlated with structural undulations and variation of membrane thickness, the inter-surface interaction between two membranes might be affected.}

A further remark should be made on the divergence of the free energy at small separations. This is related to our simplifying assumption on the boundary conditions, \eq~(\ref{eq:bc_2}), in which the electrostatic field is confined in between the surfaces, and does not leak to the outer region. Namely, the electrostatic field in the outward direction of the surface is set to zero. This assumption is reasonable when the dielectric constant of the outer region is much lower than the one of the ionic solution. However, for sufficiently small inter-surface separations, one might need to consider the leakage of the field to the outer region \cite{naji2008}.

Since the attraction in the non-linear regime was predicted only in
the limit of infinitely-large surface patches \cite{silbert2011}, it is worthwhile to investigate further the
non-linear PB model. It will be of interest to study the origin of the attraction and relate it to
the difference between the scaling of the counter-ion entropy for counter-ion release (attraction)
versus counter-ion confinement (repulsion).

Another possible extension would be to find a suitable approximation for the
 non-linear PB regime in three dimensions with heterogeneous boundary conditions.
Using such an approximation may shed more light on the interaction between
randomly charged surfaces within the PB theory, and related experiments.

{\bf Acknowledgements.~~~} We thank Henri Orland and Rudi Podgornik
for useful discussions and suggestions, and Jacob Klein, Jonathan Landy,
Sylvio May, Philip Pincus and Gilad Silbert for comments. This work was
supported in part by the U.S.-Israel Binational Science Foundation under Grant No.
2006/055 and the Israel Science Foundation under Grant No. 231/08.

\appendix*
\section{Comparison of the Linear and Non-linear PB Theory for Infinitely-large Charge Domains\label{sec:Non-linear-PN-theory}}

We repeat here the calculation presented in Ref.~\cite{silbert2011} for the non-linear PB theory and
compare it with the linear theory treated in our paper. We consider two
heterogeneous surfaces where the typical size of a charge patch (domain)
is much larger than any other length scale in the system. Each patch
can be treated as an extended section of a uniform charged surface,
and the leading interaction term would presumably be a superposition
of all patch-patch interactions between the two surfaces.

The two surfaces are overall neutral, and on each of them there are positive and negative charge patches, with uniform charge densities: $\sigma=\eta=\pm\sigma_{0}$, leading to four types of possible patch-patch interactions between the two surfaces, $\left(\pm\sigma_0,\pm\sigma_0\right)$. The electrostatic potential $\psi$ is obtained by solving numerically the
non-linear PB equation for two uniformly charged surfaces. The patch-patch interaction energy is repulsive for the like-charge case, $F_{\mathrm{rep}}>0$, and attractive for the opposite-charge case, $ F_{\mathrm{att}}<0$, and
is obtained by integrating over the
osmotic pressure $\Pi(d)$,
\begin{equation}
F/A=4\pi\kd \lb/\kbt\int^\infty_d \D z \,\Pi\left(z\right) \, ,
\end{equation}
where the osmotic pressure is given by
\begin{equation}
\Pi=-\frac{\varepsilon}{8\pi} \left(\frac{\D\psi}{\D z}\right)^2+2\kbt n_b\left[\cosh\left(\frac{e\psi}{\kbt}\right)-1\right]\, ,
\end{equation}
and has the same dependence on $\psi$ for both attractive and repulsive cases.
The overall free energy $F_{\rm av}$ is then given by averaging over the
 four possible patch-patch arrangements, yielding $ F_{\rm av}=\left( F_{\mathrm{rep}}+
 F_{\mathrm{att}} \right)/2 =\left( F_{\mathrm{rep}}- |F_{\mathrm{att}}|\right)/2$.

In \fig~\ref{fig:The-interaction-energy}(a) a comparison between
the free energy of the two cases is presented for strongly charged patches,
$\sigma=\pm\eta=10 \kd$. At  separations larger than $\kd d >0.043$,
the attractive interaction (oppositely charged surfaces, $\sigma=-\eta$, dashed line)
is stronger than the repulsive
one (equally charged surfaces, $\sigma=\eta$, solid line). This effect is related to different
scaling of the ionic entropy as a function of $d$ for the repulsive and
attractive cases. At smaller separations, $\kd d < 0.043$, the repulsive interaction
becomes stronger than the attractive one, and even diverges. The inset
shows the overall inter-surface interaction as calculated in the simplified non-linear and linear PB models. It is clear
that for large values of $\sigma$ and $\eta$ the linear approximation fails to predict the
crossover between attraction to repulsion of the non-linear PB equation.

{The origin of the difference between the linear and non-liner PB models stems from the different dependence of the counter-ions entropy on the inter-surface separation $d$ in the case of attraction and repulsion
\cite{benyaakov2007,safran_release_2007}. While the entropy in the repulsive case
is determined by counter-ion confinement \cite{andelman}, the entropy
in the attractive case is governed by counter-ion release \cite{safran_release_2007}. By directly comparing the linear and non-linear PB solution for strongly and equally-charged surfaces, it is found that the counter-ion entropy is largely overestimated in the linear PB solution~\cite{DBY_Physica_2010}. This is due to a slower decay of the ionic profile close to charged surfaces in the linear case.}

For complementarity,
in \fig~\ref{fig:The-interaction-energy}(b), we present the inter-surface interaction of weakly charged surfaces $\sigma=\pm\eta=0.5 \kd$.
As expected, the linear approximation agrees well with the simplified non-linear
PB, and the overall interaction is repulsive.
These rough estimates suggest
a possible crossover as a function of patch charge strength from repulsive to attractive inter-surface interaction,
but requires further investigations of the non-linear PB equations for surfaces with finite-size charge patches.


\end{document}